# Planetary relationships to birth (imputed conception) rates in humans: A signature of cosmic origin?

E. Georgiopoulou[1a], S. Hofmann[2], M. Maroudas[1b], A. Mastronikolis[3c], E.L. Matteson[4], M. Tsagri[1d], K. Zioutas[1*]

[1] University of Patras, Patras, Greece

[2] Untermenzig, D80999, Munich, Fastlingerstr. 17, Germany

[3] Physics Department, University of Manchester, UK

[4] Mayo Clinic College of Medicine and Science Rochester, Minnesota USA

[a] Present address: NTUA, Athens / Greece

[b] Present address: University of Hamburg, Germany

[c] Present address: CERN, Geneva, Switzerland,

[d] Present address: Rue Pierres du Niton 17, 1207 Geneva, Switzerland.

[*]Corresponding Author:

konstantin.zioutas@cern.ch

**Keywords:** conceptions, planetary relationship, dark sector, invisible matter, gravitational focusing

# Abstract

This report addresses the time dependence of normal biomedical processes. The conception rate in humans shows relationships that strikingly coincide with planetary periodicities like the orbital period of Jupiter, the 11 years solar cycle and the 237 days Jupiter- Venus synod, and the combined dependence on Jupiter with Mars' orbital position. The birth rates are used as conception surrogates, based on a data set available from the Minnesota Department of Health. The statistical significance of each independent analysis (far) exceeds $5\sigma$. This result cannot be explained with known science. As with other observations in Physics and medicine (i.e., melanoma), tentatively the only viable explanation we have is the otherwise invisible streaming matter from the dark Universe we live in. The favored dark matter scenario involves streams or clusters as opposed to an isotropic dark sector. The dark Universe scenario has been the driving idea for this type of work. The high sensitivity of living matter to the tiniest external impact might help identify the nature of the dark matter particles, a major problem in all physics since the 1930s. This work is a model for evaluating other biological processes and mutation rates.

# 1. Introduction

It has long been recognized that there are various unexplained phenomena in physics. Similarly in medicine, there are physiological patterns and diseases whose ultimate cause is unknown. Cancer is one typical example; others may include autoimmune diseases. For this work, we address the question of whether some anomalies in physics and medicine are of common origin. We focus here on planetary relationships which unexpectedly appear in both disciplines. There is no known remote planetary force that would have an observable direct effect on physical and medical processes alike. Yet, there are several investigations using regular long-term observations, that support the existence of such planetary relationships in physics and medicine [1-7].

To date, the first strong indication of a planetary origin that has engaged renowned scientists for the past ~160 years is the quasi-ubiquitous 11-year solar cycle. In order to further underpin planetary relationships [1-7], in this work we present a new 11-year periodicity in biomedicine. So far, the discussions have mainly concentrated on the extremely feeble, remote planetary effects due to gravitational tidal forces that, within typical distances between solar system bodies, cannot affect directly processes on the Sun or the Earth since they are too subtle to have any direct measurable impact at the microscopic scale. In addition, the tidal force changes smoothly during an orbit ($\sim R^{-3}$). Therefore, the occurrence of a peaking planetary relationship excludes any effect due to the omnipresent gravity, except if some specific forcing can somehow build up over a long time, which is hard to prove. For comparison with the assumptions of the present work, invisible streams, and their planetary lensing have been already discussed in the literature, with some constituents having possibly a very large cross-section σ [8-10] with normal matter (σ equal up to $\sim 10^{-24}$ cm$^2$ or even much larger). These assumptions, namely streaming DM and planetary lensing combined eventually with a large σ, make such streaming DM constituents a viable scenario for the observed planetary relationships [1-7]. In our concept, there can be a remote planetary effect on streaming DM, since its flux can be changed temporally due to gravitational focusing by the solar system bodies. Strong DM flux enhancements can occur only in the streaming scenario and not in the isotropic one.

We recall that solar and external non-solar influences on living organisms on Earth have long been the subject of scientific interest, with most publications reporting possible Lunar geotropic and light effects on plant and animal growth and reproduction [11-14]. The present work aims to assess possible planetary relationships in human reproduction based on previous observations in physics and medicine.

In the context of the present work, a publication from 1898 by Svante Arrhenius (Nobel Prize 1903) is remarkable as it addressed the question of "cosmic impact on physiological relationships" (Die Einwirkung kosmischer Einflüsse auf physiologische Verhältnisse) [15]. He was actually ahead of his time since the discovery of cosmic radiation by Hess appeared in 1912 [16]. Similarly, the underlying biology of the historic observation by Darwin in 1858 [17] remained unexplained at the fundamental level, until the discovery by Watson and Crick of the DNA double helix structure in 1953.

Of note, the dark Universe is widely assumed to consist of some generic "invisible" constituents which make up ~ 95% of its mass–energy budget. It has been discussed in the literature that DM may also exist in the form of streams or clusters or some kind of nuggets [18-22]. Their interaction with normal matter is generally expected to result in a local energy deposition of recoiling atoms or free electrons. Hence, the deposited energy could affect some of the many biochemical processes associated, for example, with pregnancy initiation, which involves a complex chain of biochemical reactions happening during conception. Then, if otherwise invisible streams from the dark sector affect some reaction in the chain of biochemical reactions occurring during cenception, the effect might be detectable as some unpredictable planetary relationship or as an "anomaly". The observed periodicities of diagnosed melanoma [4,6] reflect such anomalies.

In this work, the driving question is whether, beyond diseases, planetary links also exist for other biological events. We report here the results of an investigation aiming to examine possible effects on pregnancy initiation (derived from birth date), which are known to fluctuate according to several factors including time of year, social and lifestyle factors, and location. At first glance, the time distribution should be random and smooth, apart from seasonal or societal influences in Earth's frame of reference. We present here for the first-time planetary relationships of birth rates (i.e., imputed conception rates).

## 2. Materials and Methods

### 2.1 Data origin and curation

For this study, 3515100 births between 1970-2020 for the State of Minnesota, U.S.A, and Olmsted County were obtained from the Center for Health Statistics of the Minnesota Department of Health. For normalization, population data between 1970-2020 for the State of Minnesota and Olmsted County were obtained from the National Cancer Institute Surveillance, Epidemiology, and End Results (SEER): https://seer.cancer.gov/popdata/download.html and the United States Census (2010 to 2020): https://www2.census.gov/programs-surveys/popest/datasets/. Throughout the analysis of this work, we have used only the original raw data for the statistical analysis. Population-normalized monthly birth rates were used in establishing the apparent planetary links compared to a random Poisson distribution.

### 2.2 Possible interaction scheme

The nature of the constituents of the dark Universe and their interaction with normal matter is not yet known since ZWICKY's discovery in 1933. Observationally, we refer to generic invisible constituents from the dark sector, which might include the widely discussed DM candidates like axions and WIMPs (Weakly Interacting Massive Particles), though not the parameter phase space which is already excluded. Here we adopt from the direct DM searches the widely assumed interaction form between particles from the dark Universe and normal matter. Most of the direct DM searches, e.g., for WIMPs, are based on energy deposition by the predicted scattering of DM off normal matter, giving rise to recoil of atomic nuclei or electrons. Also, assuming that "dark photons" constitute at least part of the dark sector that is not an arbitrary assumption, converted dark photon to photon results in the emission of a real photon. This conversion can be even resonantly enhanced if the local plasma density fits the unknown rest mass of the dark photons. The expected energy depositions can be mostly locally absorbed. While there is no evidence of a direct effect of a long-suspected kind of planetary force, all solar system bodies, including the Moon, acts as gravitational lenses that can either redirect or even focus invisible streams towards the Earth, thus temporally enhancing their local flux. Hence, the various orbiting gravitational lenses in the solar system possibly impart large transient flux amplification. The influence of some solar system lenses on pregnancy initiation (derived from birth date) is presented in this work. Of note, the main signature of DM flux enhancements due to gravitational lensing is some measurement of planetary dependence of a given effect.

It is worth mentioning here, that there is still unconstrained parameter space for high cross sections (even far above ~$10^{-24}$ $cm^2$) of DM interacting with normal matter (see e.g. [8,10,21,22]) which allows such a larger interaction probability between invisible matter and normal matter, e.g., living matter. This can occur either due to a large cross-section and/or the temporally much larger fluxes. Within this construct, there is a basal flux of the same candidates from the dark Universe. It is the occasionally enhanced flux due to gravitational focusing by the solar system bodies which allows occasionally overcoming possible threshold effects. This is what makes visible the impact of the otherwise invisible streams or clusters.

Remarkably, even the gravitational self-focusing effect by the inner Earth mass distribution is expected to give rise to very large flux enhancements on the opposite side of the Earth as the incident invisible streams are exiting the Earth [23-27]. In addition, the inherent high sensitivity of living matter to external impact distinguishes it from all detectors searching for invisible matter. For example, it is known that living matter is sensitive to radioactivity also via pile-up effects, so that, for example, humans or other living organisms are a type of integrating detectors over long time intervals. This effect is mimicked occasionally by man-made detectors, at least *a posteriori* during analysis.

Since an extraordinary claim requires extraordinary evidence, we have addressed more spectral distributions. Each figure in this work shows phenomenological and model-independent evidence of a conventionally unexpected planetary relationship, with a statistical significance far above 5σ. The ultimate origin, i.e., at the microscopic level, of these observations, remains to be explained in future work. After all, the observations of Arrhenius and Darwin [15,17] would be decoded by later investigators, while the nature of ZWICKY's dark matter itself remains unknown since 1933.

In the absence of any effect, a random isotropic distribution is expected throughout the 360° of a planetary orbit. We stress throughout this work that it is the observation of an unexpected planetary relationship of any observable, be it in physics or medicine, that points to some unconventional external impact. The long-discussed planetary tidal forces hardly could come into consideration, as they are too feeble to have any direct effect within the solar system ([26] and ref.11 therein). In addition, the tidal forces are changing smoothly during one orbit, and following the often-observed peaking distributions exclude on their own a remote planetary force, which is unknown to date. Any irregularity of birth rates (imputed conception rates) in Earth's frame of reference should be smeared out in the projection on other planetary orbital positions. This is even more so when shorter orbits are involved, e.g., the inner planets. Below we present results for otherwise unexpected pregnancy initiations as they are derived from the registered birth dates. The maximum ~2 weeks uncertainty of the conception does not affect much the monthly registered birth rates.

Findings from this first-ever investigation of possible planetary connections for conceptions were obtained by applying a recently developed analysis of a long-term series of measurements of the Sun's size [7]. For Jupiter, to exclude systematics, the time interval of three complete orbits was chosen (1975-2010), yielding a striking bell-shaped distribution (see Fig. 1). Of note, we added more spectral distributions, each of which independently points at an unexpected planetary relationship.

3. Results

Since the relevant medical data are summed up monthly, for the present analysis of human conception rates (derived from birth dates) we have started with much longer planetary periodicities like the ~12 years of Jupiter (see Fig. 1) and the quasi-ubiquitous 11-year solar cycle (Fig. 2). The 11-year cycle coincides with the synod of Jupiter-Earth-Venus. I.e., a repeating synod reflects an almost identical gravitational configuration. This is important when analyzing data obtained over long-time frames since the driving concept of this type of work is based on planetary gravitational lensing effects. Encouraged by the first results from the present analysis, we also have chosen the rather short synod Jupiter-Venus of 237 days (=7.8 months). The choice is not random, but because the 7.8-month synodic rhythm is close (within 2.5%) to the 8 months as a multiple of the one-month cadence of the birth rates used throughout this work. Thus, we come to the following conclusions.

It is worth stressing here that in this work we use the registered birth rates aiming at pregnancy initiation because it is associated with several biochemical processes. If those processes are immune to the tentative external impact of this type of investigation, then any spectral shape would be random, contrary to all plots presented here. Figure 1 shows the distribution of the monthly birth rates projected on the corresponding Jupiter heliocentric longitudes, and this is for two different BIN sizes. During the 51-year time interval of this study, only three consecutive Jupiter orbits are available when starting at zero heliocentric longitude. This choice suppresses possible artifacts. As an example, if one would start at any other orbital point, a step can show up as the existing conditions can not be identical over large time intervals. Both plots show dependency on the orbital position of Jupiter, with the bell-shaped Figure 1b being well pronounced. Using only the two values of its minimum and its maximum (Fig. 1b), the estimated amplitude is equal to (7.9 ± 0.3) %. The modulation amplitude of birth rates (imputed conception rates) in Jupiter's frame of reference has a statistical significance far above 5σ (=7.9/0.3 = 26 σ). The statistical significance is given by the underlying number of conceptions as being derived from the birth dates. Of note, in Figure 1b each bin value is based on about 250000 pregnancy initiations, i.e., with a statistical accuracy of 0.2%, or, to put it differently, the accuracy of the amplitude (=7.9%) based on the difference of only the two values of maximum – minimum is 0.28%. The high significance is evident even on visual inspection. This trend in Figure 1b along Jupiter's orbital longitudes is also seen in Figure 1a with the smaller cadence (BIN=10º). Interestingly, the observed 12 peaks reflect the 12 annual modulations in Earth's reference frame, which we neglect in order to suppress seasonal or other systematics as a possible planetary dependency of human conception rates (following the registered month of birth).

Further, an important question is whether birth rates follow the 11-year solar cycle. Figure 2 confirms this relationship with a significance far above 5σ. Interestingly, using the original data without a population normalization of the birth rates gives a different shape (not shown here) with a similar amplitude (~13%). This indicates that a robust 11-year periodicity is at work.

Attention is also given to the behavior of a synod of two planets. Namely, the Venus-Jupiter synod is chosen because the time interval of 8 months is close to the synod Jupiter-Venus (7.8 months). The non-random spectral shape shown in Figure 3 is apparent and highly significant (far above 5σ), and this is even while adding 76 consecutive synods (Figure 3 *left*). The summing up of 76× should diminish any possible artifact. In addition, by splitting the whole range into two halves, the shape remains strikingly similar in all three plots of Figure 3. This splitting excludes even some unforeseen systematics.

## 4. Discussion

The possibility of exo-Earth influences on birth rates is not new and was most recently examined in relationship to the lunar cycle [28]. The driving idea behind this investigation was that some form of invisible streaming matter from the dark Universe could affect biological processes. The present work provides evidence based upon several lines of analysis showing that the studied birth rates are related to various planetary configurations, from simple planets to combined planets, e.g., synods. Interestingly, the birth rates also follow the solar cycle of 11 years, which was suspected to be planetary in origin as early as 1859 [29]. As already argued in ref. [1], some generic invisible streams from the dark Universe could also influence such events as pregnancy initiation, as reflected in the present observations for which we have no other viable explanation. If invisible non-relativistic streams or clusters are involved, then both cases can result in a temporal flux enhancement due to planetary gravitational lensing avoiding constraints due to threshold effects.

Living matter has an inherent high sensitivity to external irradiation, and the generic invisible streams or clusters are an as-yet overlooked component of low-speed cosmic radiation. We recall that the same invisible scenarios have motivated physics investigations [2,3] and have been used as a possible explanation for the behavior of many observables [7, 26]. The 11-year solar cycle also reflects planetary relationships following several physics observables. It has been proposed that the cause must be due to some generic streaming invisible massive matter and that when a low-speed stream is aligned toward the Earth with an intervening planet or the Sun, or even the Moon, its influx increases temporally due to gravitational focusing, assisted eventually by the self-focusing effect of the inner Earth's mass distribution. The flux amplification factor due to gravitational focusing within the solar system can be several orders of magnitude, occasionally strongly modifying the influx of invisible matter, for example, towards the Earth or any other solar system body (for more details about this methodology see [27]).

We still can only speculate about the invisible matter scenario, that could be at the origin of various unexplained phenomena in the solar system. Notably, a conventionally unexpected planetary relationship for some observable(s) is the novel key signature in favor of "invisible" streaming matter. Such new out-of-the-box signatures are being investigated in this study.

Here we present planetary relationships for the time-dependent monthly birth rates in a geographically defined population by projecting their values to the planetary heliocentric longitudes (see Figure 1 and Figure C1 in the complementary material). In the future, using daily data which implies a better time resolution, rates influenced by the inner planets including the Moon could be examined. This work should also be seen as accumulating evidence for the underlying process(es), which are of importance for deciphering the underlying physics which are still speculated about. We propose that future epidemiologic *in vivo* and also *in vitro* investigations be undertaken with other observables in bio-medicine, potentially including mutation rates as they may test and validate these first observations with an in depth analysis.

## 5. Conclusion

As previously noted for melanoma occurrence [4,6], a planetary relationship is the novel signature for some external otherwise invisible impact on biological processes related to conception as reflected in birth rates. Streams from the dark Universe are the only viable explanation we have for the observed effects.

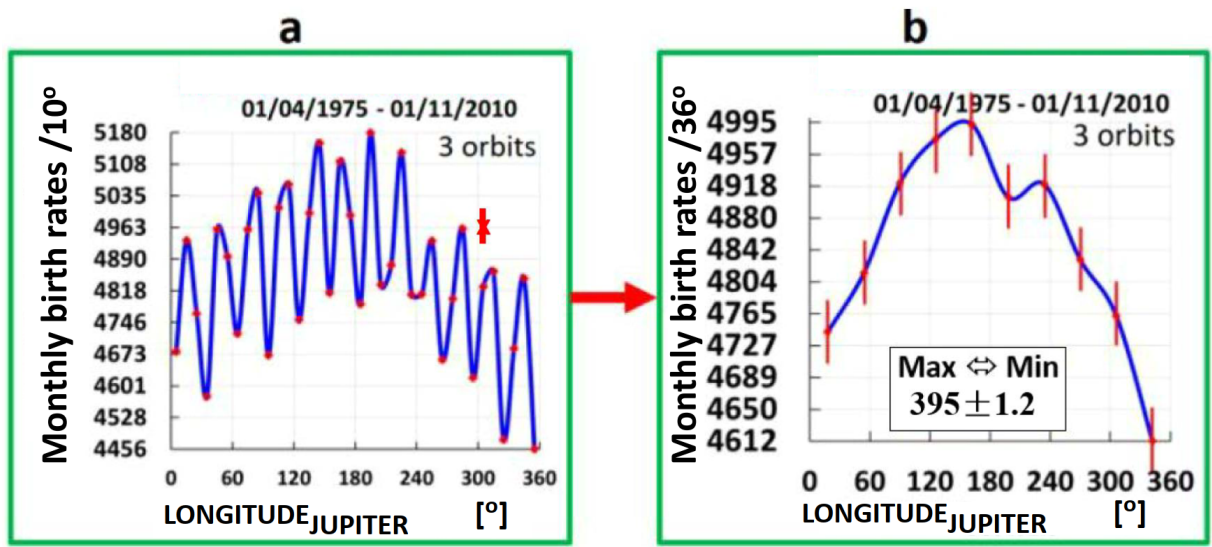

**Figure 1:** Population normalized birth rates for 3 complete Jupiter orbits, 1975-2010, without planetary constraints. (**a**) For BIN=10º, the reproduced 12 annual oscillations during the ~12-year orbital period of Jupiter support the credibility of the applied analysis. The distribution has a maximum around the heliocentric longitude of ~180º. The error bar (in red) is equal to ±2σ. (**b**) Using BIN=36º, the associated averaging gives a much smoother distribution than the left plot. The estimated relative difference maximum-minimum of ~ (7.9±0.3)% is based only on the corresponding two extreme count rates; assuming Poisson distribution, it yields a statistical significance for the amplitude far above 5σ (=7.9/0.3 = 26σ). The error bars (in red) are equal to ±3.8σ. The underlying number of births is about 250000/BIN. To exclude possible artifacts, the three selected consecutive Jupiter orbits start at zero heliocentric longitude.

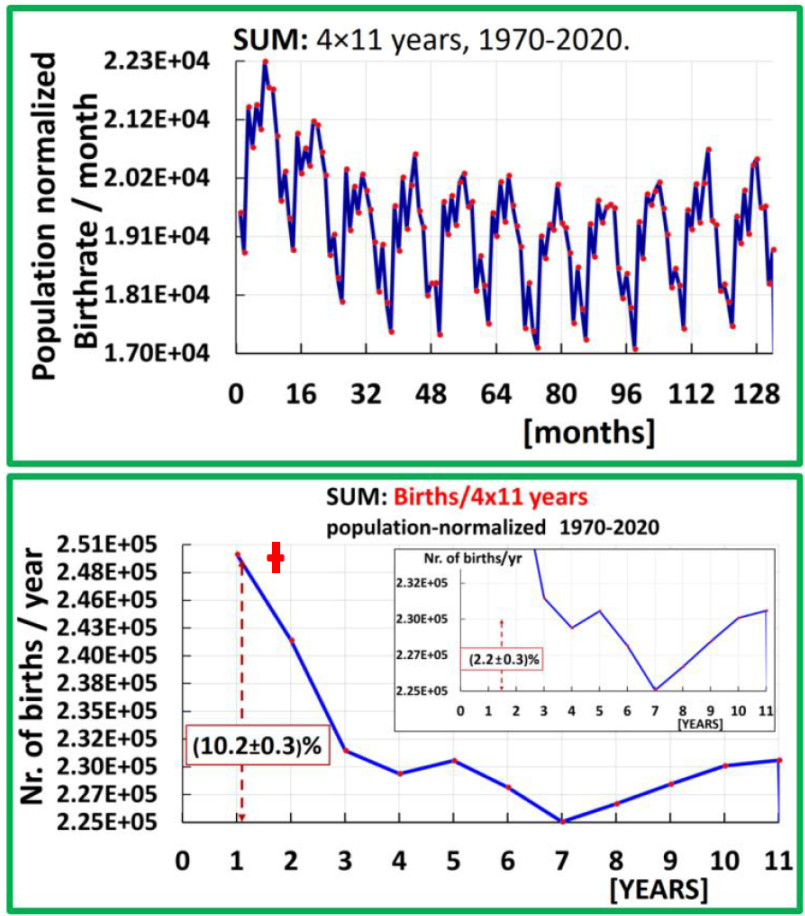

**Figure 2:** *Upper panel:* The summing-up of four times 132 consecutive monthly birth rates (=4× the 11 years solar cycle) starting January 1970. The maximum–minimum difference is 23.4%. Assuming Poisson statistics, the mean standard deviation is σ≈0.7%. The apparent annual modulation repeats 11 times being also supportive of the present analysis.

*Lower panel:* The same distribution is shown with a much larger integrating time interval (BIN=1 year). The amplitude is (10.2 ± 0.3)% with a large statistical significance (>30σ). The error bar (in red) is equal to ±3σ. The region around the minimum at year 7 is shown expanded in the *insert*: comparing the dip in year 7 to the next two points on either side has a significance of above 5σ. The sum of the monthly birth rates is population normalized.

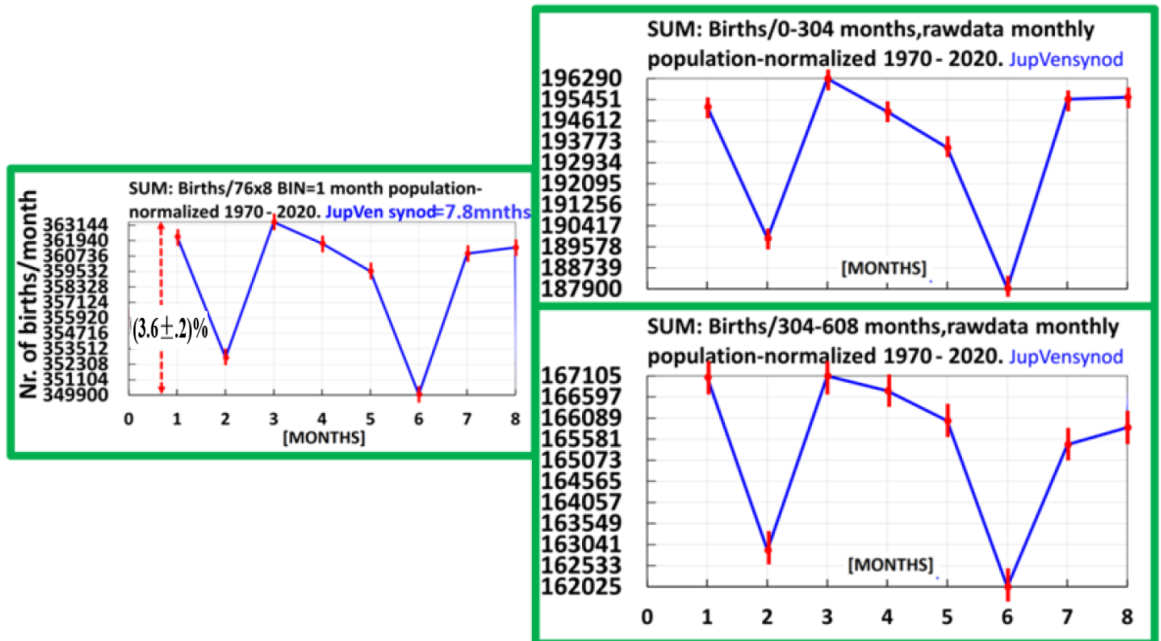

**Figure 3:** The 237 days Jupiter-Venus synod (= 7.8 months). *Left:* The complete time interval 1970-2020. The statistical significance only for the dip at 6 months, assuming Poisson distribution, is far above 5σ. *Right:* The spectral shape remains the same when using 2× half the time interval used on the *left*, or 2×304 months. The striking similarity between the two halves and that of the total distribution on the *left* excludes systematics. The monthly birth rates used are population normalized. The standard deviation for each of the 8 points is ~ 0.17% (*left* plot) and ~ 0.24% (*right* plots).

**Data availability:**
The data supporting the findings of this study are available from the corresponding author upon reasonable request.

**Author contributions:**
Conceptualization: KZ
Proposal: EG
Resources: ELM, KZ
Data curation: ELM, KZ, MM
Methodology: ALL
Formal Analysis: KZ, MM, AM
Validation: ALL
Writing – original draft: ALL
Writing – review & editing: ALL

**Competing interests:**
The authors declare no competing interests.

**Ethics approval and consent to participate:**
Ethics permission was not applied for, as all birthrate data underlying the study are publicly available. No patient information was accessed and only deidentified summary statistics were used for birth rates. Hence, an ethics committee evaluation was not needed according to the guidelines.

**Correspondence** and requests for materials should be addressed to Konstantin Zioutas.

# COMPLEMENTARY MATERIAL

The spectral shape in the reference frame of Jupiter either without any constraint (Figure 1) or with a planetary constraint imposed on the heliocentric orbital arc where Mars is propagating is shown in Fig. C1. Here the entire birth rate dataset 1970-2020 is selected. Again, in the absence of an external impact, all spectral shapes in this work should be isotropic and randomly distributed. None of the plots of Fig. C1 exhibit a random distribution; the statistical significance of the difference between the maximum-minimum points is above 5σ. The apparent different spectral shape of each of the three plots strengthens the evidence for the presence of a statistically significant peaking planetary relationship in the distribution of birth rates, and, by extension, the conception rate.

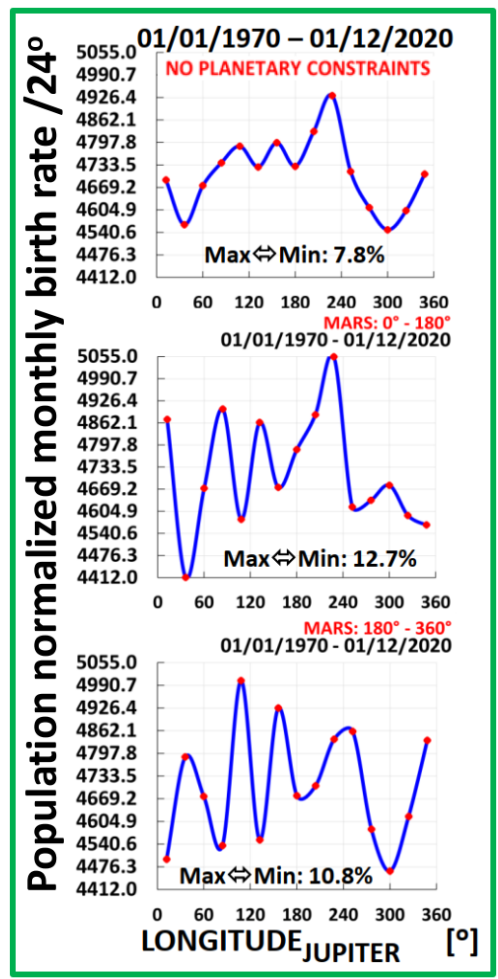

**Figure C1:** Population normalized birth rates as a function of Jupiter's heliocentric orbital position (BIN=24º). *Top:* without any planetary constraints. *Middle & bottom:* combined with orbital constraints of the planet Mars. The relative maximum-minimum difference is obtained by using only the two extreme points in each panel. From top to bottom panel, the mean standard deviation per point is about 0.2%, 0.29%, and 0.3%. With the amplitudes of the difference between the maximum and minimum point of each panel, the statistical significance for the peaking distributions is far above 5σ for all three plots.